\documentstyle{crnart12}
\def\gappeq{\mathrel{\rlap {\raise.5ex\hbox{$>$}}
{\lower.5ex\hbox{$\sim$}}}}

\def\lappeq{\mathrel{\rlap{\raise.5ex\hbox{$<$}}
{\lower.5ex\hbox{$\sim$}}}}

\def\beq{\begin{equation}}
\def\eeq{\end{equation}}
\def\bea{\begin{eqnarray}}
\def\eea{\end{eqnarray}}
\def\PL{{\it Phys.Lett.} }
\def\PRL{{\it Phys.Rev.Lett.} }
\def\NP{{\it Nucl.Phys.} }
\def\PR{{\it Phys.Rev.} }

\def\ZP{{\it Z.Phys.} }

\def\Toprel#1\over#2{\mathrel{\mathop{#2}\limits^{#1}}}
\def\Pbar{\Toprel{\lower 2pt
\hbox{$\scriptscriptstyle(-)$}}\over{\partial}}

\def\Toprel#1\over#2{\mathrel{\mathop{#2}\limits^{#1}}}
\def\Gbar{\Toprel{\lower 2pt
\hbox{$\scriptscriptstyle(-)$}}\over{G}}

\def\Toprel#1\over#2{\mathrel{\mathop{#2}\limits^{#1}}}
\def\pbar{\Toprel{\lower 2pt
\hbox{$\scriptscriptstyle(-)$}}\over{P}}

\def\Toprel#1\over#2{\mathrel{\mathop{#2}\limits^{#1}}}
\def\Pibar{\Toprel{\lower 2pt
\hbox{$\scriptscriptstyle(-)$}}\over{\Pi}}

\begin{document}
\pagestyle{empty}
\begin{flushright}
{CERN-TH.6943/93}
\end{flushright}
\vspace*{1.5mm}
\begin{center}
{\bf DO PRESENT LEP DATA PROVIDE EVIDENCE}\\
{\bf  FOR ELECTROWEAK CORRECTIONS?} \\
\vspace*{1cm}
{\bf  V.A. Novikov} \\
\vspace*{0.3cm}
ITEP, Moscow 117259, Russia \\
\vspace*{0.5cm}
{\bf L.B. Okun}$^{*)}$ \\
\vspace*{0.3cm}
 Theoretical Physics Division, CERN\\
 CH-1211 Geneva 23, Switzerland \\
\vspace*{0.5cm}
and \\
\vspace*{0.5cm}
 {\bf M.I. Vysotsky}\\
\vspace*{0.3CM}
 ITEP, Moscow 117259, Russia\\
 \vspace*{1.5cm}
{\bf ABSTRACT}
\vspace*{0.2cm}
 \end{center}
The Born approximation, based on $\bar\alpha \equiv\alpha (m_Z)$ instead
of
$\alpha$, reproduces all electroweak precision measurements within their
$(1\sigma)$ accuracy. The low upper limits for the genuinely
electroweak
corrections constitute one of the major achievements of LEP. The
astonishing
smallness of these corrections results from the cancellation of a large
positive contribution from the heavy top quark and large negative
contributions
from all other virtual particles. It is precisely the non-observation of
electroweak radiative corrections that places stringent upper and lower
limits
on the top mass.

\vspace*{3cm}
\noindent
\rule[.1in]{16.5cm}{.002in}

\noindent
$^{*)}$ Permanent address: ITEP, Moscow 117259, Russia.

\begin{flushleft} CERN-TH.6943/93 \\
July 1993
\end{flushleft}

\vfill\eject
\pagestyle{empty}

\setcounter{page}{1}
\pagestyle{plain}

The precision measurements of $Z$ decays at LEP are usually considered
as
providing evidence for  non-vanishing electroweak radiative corrections
(see e.g.
\cite{aaa}, where a representative list of references is given). The aim
of this
letter is to stress  that present LEP data \cite{bb}-\cite{ee} are in
perfect
agreement \cite{ff}-\cite{hh} with the Born approximation and that no
genuine
electroweak corrections (involving loops with heavy virtual bosons,
neutrinos
and top quarks) have  as yet been observed.  The disagreement with
 statements to the contrary stems from the  different definitions of the
Born approximation being used. Usually it is defined in terms of
electric charge
at zero momentum transfer, i.e.
\begin{equation} \alpha \equiv \alpha(0) = e^2/4\pi =
1/137.035 9895(61) \;\; ,
\label{1}
\end{equation}
while we
argue that the true Born approximation should be defined in terms of
\begin{equation}
\bar{\alpha} =
\alpha(m_Z) = 1/128.87(12)\;,\quad\quad {\rm [9],[10]}~.
\label{2}
\end{equation}
By using
$\bar{\alpha}$  instead of $\alpha$ one automatically takes into account
the only purely electromagnetic correction (polarization of vacuum by
the
photon), which has not already been allowed for
by the experimentalists.

While $\alpha(q^2)$ is running, the  other two gauge couplings
\begin{equation}
\alpha_W = g^2/4\pi \;\;, \;\; \alpha_Z = f^2/4\pi
\label{3}
\end{equation}
are ``frozen" for $\vert q^2 \vert \leq m^2_{Z,W}$ and start to run only
for
$\vert q^2 \vert \gg m^2_{Z,W}$. Therefore it is natural to consider the
electroweak  Born approximation at the Fermi scale, i.e. at $q^2 \simeq
m^2_Z$.
In a sense, $\alpha$, with all its accuracy, is irrelevant to
electroweak
physics; what is relevant is $\bar{\alpha}$. Hence, if  Glashow,
Weinberg and
Salam \cite{lll}-\cite{nn} had thought about actually calculating
electroweak radiative corrections
they would have used $\bar{\alpha}$ from the beginning. Then they would
have
defined the weak angle $\theta$ through the equations
\begin{equation}
\alpha_W = {\alpha}_Z c^2\;\;, \; \bar{\alpha} = \alpha_W s^2\;\;,
\label{4}
\end{equation}
where $c\equiv \cos\theta$, $s\equiv\sin\theta$. (We do $\underline{\rm
not}$ use
$\theta_W, \; s_W, \; c_W$ here, because in the literature they are
associated
with $\alpha$, not $\bar{\alpha}$.) Hence
\begin{equation}
c^2 s^2 =
\bar{\alpha}/\alpha_Z \;\; .
\label{5}
\end{equation}
According to the
Minimal Standard Model \cite{lll}-\cite{nn}:
\begin{equation}
m_W = g\eta/2 \;\; , \;\; m_Z =
f\eta/2
\label{6}
\end{equation}
where $\eta$ is the vacuum expectation value of the Higgs field,
so that in the Born approximation:
\begin{equation}
m_W/m_Z=c \;\; .
\label{7}
\end{equation}

To obtain $\eta$ we consider, as usual, the
four-fermion coupling of  $\mu$-decay (see, for instance, \cite{oo})
\begin{equation}
\frac{G_{\mu}}{\sqrt{2}}=\frac{g^2}{8m^2_W} \;\; .
\label{8}
\end{equation}
Then it follows from (\ref{6}):
\begin{equation}
\eta^2 = 1/\sqrt{2} G_{\mu} \;\; .
\label{9}
\end{equation}
The value
\begin{equation}
G_{\mu}=1.16639(2)\times 10^{-5}~ {\rm GeV}^{-2}
\label{10}
\end{equation}
gives:
\begin{equation}
\eta = 246.2185(21)~{\rm GeV} \;\; .
\label{11}
\end{equation}

In the pre-LEP era, both $m_Z$ and $m_W$ were poorly known. This
justifies
the definition \cite{pp} $s_W\equiv m_W/m_Z$ from the historical point
of view.
At present, however, $m_Z$ is known with much higher accuracy than $m_W$
\cite{qq}:
\beq
m_Z = 91.187(7)~{\rm GeV}
\label{12}
\eeq
\beq
m_W = 80.22(26)~{\rm GeV}
\label{13}
\eeq
 It is
reasonable therefore to express $s$ and $c$ in terms of $m_Z$:
\begin{equation}
f^2 = 4m^2_Z/\eta^2=4\sqrt{2} G_{\mu}m^2_Z = 0.548636(84)
\label{14}
\end{equation}
\begin{equation}
\alpha_Z = \frac{\sqrt{2}}{\pi}\cdot G_{\mu}m^2_Z = 1/22.9047(35) =
0.0436592(66)
\label{15}
\end{equation}
\begin{equation}
\frac{1}{4}\sin^2 2\theta = c^2 s^2 =\bar{\alpha}/\alpha_Z =
\frac{\pi\bar{\alpha}}{\sqrt{2}G_{\mu}m^2_Z}= 0.177735(16)
\label{16}
\end{equation}
\begin{equation}
s= 0.48081(33) \;\;, \;\; c= 0.87682(19)
\label{17}
\end{equation}
\begin{equation}
s^2 = 0.23118(33) \;\;, \;\;  c^2 = 0.76881(33)~.
\label{18}
\end{equation}
Now we are ready to derive the electroweak Born predictions for various
observables. We first compare Eq. (\ref{7}) with the experimental ratio
\beq
m_W/m_Z = 0.8797(29)~.
\label{19}
\eeq
The agreement is within $1\sigma$.
Next we  consider decays of the $Z$-boson. The amplitude of the
decay into pairs of charged leptons, $e^+ e^-, \; \mu^+\mu^-, \;
\tau^+\tau^-$, has the form:
\begin{equation}
M_l = \frac{1}{2}f\bar{l}[g_A\gamma_{\alpha}\gamma_5 +
g_V\gamma_{\alpha}]l Z_{\alpha}~,
\label{20}
\end{equation}
where $l$ and $Z_{\alpha}$ are the wave functions of lepton and
$Z$-boson.

The corresponding width $\Gamma_l$ is given by the expression:
\begin{equation}
\Gamma_l = \bigg(1+\frac{3\bar{\alpha}}{4\pi}\bigg)\times 4(g^2_A +
g^2_V)\Gamma_0~,
\label{21}
\end{equation}
where
\begin{equation}
\Gamma_0=\frac{f^2 m_Z}{196\pi}=\frac{\sqrt{2}G_{\mu}m^3_Z}{48\pi}=
82.941(19)~{\rm MeV}~.
\label{22}
\end{equation}

The forward-backward asymmetry in the channel $f\bar{f}$ is
\begin{equation}
A_{FB}=\frac{3}{4}A_e A_f \;\; ,
\label{23}
\end{equation}
where
\begin{equation}
A_i = 2g^i_A g^i_V/(g^{i^2}_A + g^{i^2}_V)~,\quad (i=e,\mu \ldots )
\label{24}
\end{equation}
and the longitudinal polarization of the $\tau$-leptons
\begin{equation}
P_{\tau}=-A_{\tau}~.
\label{25}
\end{equation}
As is well known, the Born aproximation gives, for charged leptons
\begin{equation}
g_A = T_3 = -1/2 = -0.5000~,
\label{26}
\end{equation}
\begin{equation}
g_V/g_A = 1-4s^2 = 0.0753(12)~,
\label{27}
\end{equation}
which should be compared with corresponding experimental values
\cite{bb}-\cite{ee}:
\begin{equation}
g_A^{exp}=-0.4999(9)
\label{28}
\end{equation}
and
\begin{equation}
(g_V/g_A)^{exp}=0.0728(28) \;\; ,
\label{29}
\end{equation}
(the latter was obtained from the measurements of $A_{FB}$ for leptons
and
hadrons, and from $\tau$-polarization). Again we see agreement to within
$1\sigma$.

The decays into hadrons may be considered as decays into quark +
antiquark
pairs. In this case, as before,
\begin{equation}
g_A = T_3 \;\; ,
\label{30}
\end{equation}
but the  fractional charges and the colour degrees of freedom of the
quarks must
be taken into account:
\begin{equation}
g_V/g_A = 1-4\vert Q\vert s^2~,
\label{31}
\end{equation}
\begin{equation}
\Gamma_q = 12\bigg(1+\frac{3}{4\pi}Q^2\bar{\alpha}\bigg)G\Gamma_0\times
(g^2_A + g^2_V) \;\; ,
\label{32}
\end{equation}
where the factor $G$ describes the final state exchange and emission of
gluons
\cite{rr}-\cite{tt}
\begin{equation}
G=1+\bar{\alpha}_s/\pi + 1.4(\bar{\alpha}_s/\pi)^2 -
13(\bar{\alpha}_s/\pi)^3 + ...
\label{33}
\end{equation}
Here $\bar{\alpha}_s\equiv\alpha_s(m_Z)$ is the   gluonic
coupling constant at the Fermi scale.
For further estimates we will assume that
\begin{equation}
\bar{\alpha}_s = 0.12\pm 0.01 \;\; ,
\label{34}
\end{equation}
which agrees with  the global analysis of all
pertinent data: $\bar{\alpha}_s = 0.118 \pm 0.007$ \cite{uu},\cite{vv}.
Then
\begin{equation}
G(\bar{\alpha}_s = 0.12 \pm 0.01) = 1.0395(33)
\label{35}
\end{equation}
LEP data are compared with the Minimal  Standard Model predictions in
the Table.

Note that in the Table the ``Born" values of hadronic observables are
obtained
by using the gluonic factor $G$, given by Eq. (33) for all quark
flavours. The
specific gluonic corrections to $\Gamma_b$ \cite{vv}-\cite{cci} caused
by the
non-vanishing $\bar m_b = m_b(m_Z) = 2.3$ GeV and the large $m_t$ are
included
in the MSM corrections. Allowing for them in the $G$-factor of $b\bar b$
decay
would give $G_B \simeq G - 0.01$; the new central Born values ($\Gamma_h
1739$ MeV, $\Gamma_Z$ = 2487 MeV, $\sigma$ = 41.46 nb, $R_{\ell}$ =
20.82 and
$R_b$ = 0.218) would still preserve the $1\sigma$ agreement with
experimental
data.

The agreement between ``Born" and experiment is stunning, even if one
allows for
the fact that not all the observables in the Table  are independent:
$\Gamma_{\ell}$ can be expressed in terms of $g_A$ and $g_V/g_A$, $R_l =
\Gamma_h/\Gamma_{\ell}$ and
\begin{equation} \sigma_h =
12\pi\Gamma_e\Gamma_h/m^2_Z\Gamma^2_Z ~.
\label{36}
\end{equation}
The coincidence of the central experimental and Born
values is amazing: their differences are in some cases  smaller than the
experimental uncertainties. This fact must  be considered  a rare
statistical fluctuation.

What is much more interesting from the physics point of view is the
smallness of the electroweak radiative corrections as compared with
na\"\i ve
estimates $(\sim\alpha_W/\pi\sim\bar\alpha )$. This  originates from the
compensation of two large contributions: a positive one from the heavy
top quark
($m_t\sim 150$ GeV), and a negative one from all other virtual particles
(light
quarks, higgs, $W,Z$-bosons).

 Were the top quark much
lighter, the agreement with the Born approximation would be destroyed.
This is
shown in detail in \cite{ggg},\cite{hh}.
It may seem paradoxical, but it is precisely the non-observation of
electroweak
radiative corrections that places stringent upper and lower limits on
the
top's mass.
(Note that in the usual approach based on $\alpha$, not $\bar{\alpha}$,
the
same limits appear as a result of precision measurements of
non-vanishing
radiative corrections, some of which are very large.)

The results of the ``low-energy" electroweak experiments complement
the above
picture of ``Born"--experiment agreement. From $\nu_{\mu}e$ scattering
experiment \cite{ddi}:
\begin{equation}
g_A^{e\nu_{\mu}}= -0.5030(180) \; , \;\;\;
g_V^{e\nu_{\mu}}/g_A^{e\nu_{\mu}}=
0.0500(380) \;\; ,
\label{37}
\end{equation}
which should be compared with $-0.5000$ and $0.0753(12)$, respectively
[29].
The three experiments on deep inelastic neutrino scattering  (CHARM,
CDHS, CCFR)
give for $m_W/m_Z = 0.8785(30)$ (see \cite{qq}), where the uncertainty
seems
to be less reliable than in the case of direct $m_W$ measurement (UA2,
CDF).
Still, it would be interesting to check whether the electroweak Born
approximation (with due allowance for strong interactions) would
describe
deep inelastic scattering also to within $1\sigma$. The most promising
seems to
be the experiment  on atomic parity violation in $~^{133}$Cs. The
experimental
value of the weak charge $Q_W(^{133}$Cs)=-71.04(1.81) is $1.5\sigma$
away from
its ``Born" value: $-73.9$.

The reduction of the one-loop corrections by a factor 3 to 5 (through
partial
cancellation) is important in the light of the experimental
uncertainties. Even
if these were reduced by, say, a factor of 3,  this would still not make
the
electroweak two-loop contribution essential. Hence one can safely limit
oneself
to the one-loop electroweak approximation.

The fact that one can confine oneself from the beginning to the one-loop
approximation not only makes the
calculations quite simple but also  extremely transparent. It is
convenient
to organize them in five steps:
 \begin{description}
\item[Step 1.] Start with a Lagrangian, which contains only bare
couplings
$(e_0, f_0, g_0)$ and bare masses $(m_{W0},m_{Z0}\ldots)$. Substitute
$m_t,
m_b$ and $m_H$ for
$m_{t0}, m_{b0}$, and $m_{H0}$,  since two loops are
neglected. \item[Step 2.] Calculate one-loop Feynman diagrams for the
three most
accurately known observables -- $G_{\mu}, m_Z, \bar{\alpha}$ -- in terms
of the bare quantities $e_0, f_0$ ... and $1/\varepsilon$,
where $\varepsilon$ is the parameter used in  dimensional
regularization:
$2\varepsilon = D-4$ and $D$ is the dimension in which Feynman integrals
are
calculated.
\item[Step 3.] Invert the   equations resulting from Step 2 by
expressing all
bare quantities in terms of $G_{\mu}, m_Z, \bar{\alpha}~, m_t, m_b, m_H$
and $1/\varepsilon$.
\item[Step 4.] Calculate Feynman integrals for $m_W/m_Z, g_A, g_V/g_A$
or
any other electroweak observable in terms of bare quantities $e_0, g_0$
...
and $1/\varepsilon$.
\item[Step 5.] Express $m_W, g_A, g_W/g_A$, etc, in terms of
$G_{\mu}, m_Z, \bar{\alpha}, m_t, m_b, m_H$. At this step all terms
proportional to $1/\varepsilon$ cancel each other and the resulting
relations
contain no infinities in the limit $\varepsilon \rightarrow 0$.
\end{description}

Each of the ``gluon-free" observables, $m_W/m_Z, g_A, g_V/g_A$, is
conveniently  presented as the sum of the Born term and the one-loop
term
\cite{ggg}:
\begin{eqnarray}
m_W /m_Z &=& c+\bar{\alpha}\frac{3c}{32\pi s^2(c^2 -s^2)}V_m(t,h)=
\nonumber
\\
&=&0.8768 + 0.00163 V_m~,
\label{38}
\end{eqnarray}
\begin{eqnarray}
g_A &=& -\frac{1}{2}-\bar{\alpha}\frac{3}{64\pi s^2 c^2}
V_A(t,h)= \nonumber \\
&=&-0.5000 - 0.00065 V_A~,
\label{39}
\end{eqnarray}
\begin{eqnarray}
g_V/g_A &=& 1-4s^2 +\bar{\alpha}\frac{3}{4\pi(c^2- s^2)}
V_R(t,h)= \nonumber \\
&=&0.0753(12) + 0.00345 V_R~,
\label{40}
\end{eqnarray}
where
\begin{equation}
t=(m_t/m_Z)^2 \;\; , \;\; h=(m_H/m_Z)^2~.
\label{41}
\end{equation}
All three functions $V_i$ are normalized in such a way that
they behave similarly for $t\gg 1$, i.e.:
$$ V_i\simeq t \;\; \mbox{for}~ t\gg 1 \; .$$

By comparing Eqs. (\ref{38}), (\ref{39}), (\ref{40}) with the
corresponding
experimental values (see column 1 of the Table ) one  obtains the
experimental values of the $V_i$'s:
\begin{equation}
V_i^{exp}=\bar{V}_i\pm \delta V_i~.
\label{42}
\end{equation}
They are:
\beq
V_m^{\exp} = 1.78 \pm 1.78
\label{43}
\eeq
\beq
V_A^{\exp} = -0.15 \pm 1.38
\label{44}
\eeq
\beq
V_R^{\exp} = -0.73 \pm 0.81~.
\label{45}
\eeq
The fact that experiments are, to within $1\sigma$,  described by the
Born
approximation means that
\begin{equation}
\vert \bar V_i\vert \leq\vert\delta V_i\vert~.
\label{46}
\end{equation}
The one-loop approximation leads to an important property of the
functions
$V_i(t,h)$, namely, they may be presented in the form \cite{ggg}:
\begin{equation}
V_i(t,h)=t+T_i(t)+H_i(h) \;\; .
\label{47}
\end{equation}
Thus $V_i(m_t)$ for different values of $m_H$ differ only by  a
shift; see Fig. 1, where $V_R(m_t)$ is presented.

Simple analytical expressions and numerical tables for functions
$T_i(t)$ and
$H_i(h)$ are given in Ref. \cite{ggg}, which allows a ``do-it-yourself
analysis" [30] of the data. It is important to emphasize that Eqs.
(38)-(40) are
$\underline{\rm exact}$ in the one-loop approximation, unlike the
so-called
``improved Born approximation" (see, e.g., [31]), which starts with
$\alpha$  and includes terms proportional to $\Delta\alpha = \bar\alpha
\alpha$ and $t$.

The functions $V_i(t,h)$ form a surface over the plane $m_t, m_H$. The
intersection with a plane orthogonal to the axis $m_H$ gives a curve
describing
the $m_t$-dependence of $V_i$ at given $m_H$. Examples of such curves
are given
in Fig. 1. Similarly, the intersection with a plane orthogonal to the
axis $m_t$
gives a curve describing the $m_H$-dependence of $V_i$ at given $m_t$.
Horizontal planes at $V_i = \bar{V}_i \pm\delta V_i$ give isolines
corresponding
to a central value $\bar V$ and $1\sigma$ uncertainties (see Fig. 2).
The
crossing point of central-value isolines  determines in principle the
values of
$m_t$ and $m_H$. Unfortunately the
 $\delta V_i$'s are  so large that reliable limits may be obtained
only for $m_t$.

As for $m_H$, the minimum $\chi^2$ lies  at $m_H =$ 10 GeV, which is
much below the lower experimental LEP bound (62 GeV). This false minimum
corresponds to the crossing point of central-value isolines in Fig. 2.
It is
evident that this contradiction is statistically not significant: by
shifting
the $V_A$ isolines to the right by slightly more than $1\sigma$ one can
readily
get the crossing point at an $m_H$ of several hundred GeV.

A similar analysis
may be performed for the hadronic decays of $Z$-bosons. As basic
observables
one can choose  $\Gamma_Z$, $R_l=\Gamma_h/\Gamma_l$ and $\sigma_H$,
which
depend on $\bar{\alpha}_s$, and also $R_b$, whose dependence is less
pronounced.
{}From these observables, stringent limits can be obtained not only on
$m_t$, but
also on $\bar\alpha_s$ \cite{hh}. Our results for $m_t$ and
$\bar\alpha_s$ are
in qualitative agreement with the results of $\chi^2$ fits published by
other
authors (see, for instance: \cite{aaa},\cite{dd}, \cite{qq},[32],
[33]).

In the last three columns of the Table, we give the values of MSM
radiative
corrections, which illustrate the sensitivity of various observables to
the
values of $\bar\alpha, \bar\alpha_s$ and $m_t$. (These corrections also
include
the effects of virtual gluons in electroweak quark loops [34]).

When the data from LEP and the SLC have a better accuracy, and when the
top
quark is discovered and its mass is measured, the MSM corrections may
become
non-adequate. This would signal the existence of New Physics. A
convenient
parametrization of it has already been worked out [35].

\vspace*{1cm}
\noindent
{\bf ACKNOWLEDGEMENTS}

We are grateful to G. Al\-ta\-rel\-li, R. Bar\-bi\-eri, D. Bardin,
J. El\-lis, E. Li\-si, A.~Olchevski, A. Rozanov, V. Te\-leg\-di and V.
Yurov for many
interesting discussions.  We are grateful to the Russian Foundation for
Fundamental Research for grant Nr. 93-02-14431 supporting this work. One
of us
(LO) thanks the CERN Theory Division for their warm hospitality.
\vfill\eject

\begin{table}
\begin{center}
\begin{tabular}{|l|l|l|l|l|l|l|l|} \hline
\multicolumn{1}{|c|}{1} & \multicolumn{1}{c|}{2}  &
\multicolumn{1}{c|}{3}  & \multicolumn{1}{c|}{4}  &
\multicolumn{1}{c|}{5}  & \multicolumn{1}{c|}{6}
& \multicolumn{1}{c|}{7}  & \multicolumn{1}{c|}{8}\\ \hline
Observable &
Exp. value & \multicolumn{1}{c|}{``Born"} & \multicolumn{3}{c|}{MSM
corrections} & \multicolumn{2}{c|}{$\Delta m_t$ shifts  }\\
\hline
$m_W/m_Z$ & ~0.8798(29) & ~0.8768(2) & ~33 & ~25 & ~17  & 7 & 13\\
$g_A$ & -0.4999(9) & -0.5000 & -8 & -7 & -4 & $\underline{2}$ &
$\underline{4}$\\
$g_V/g_A$ & ~0.0728(28) & ~0.0753(12) & -31 & -55 & -75 & 12 & 25\\
$\Gamma_{\ell}$ (MeV) & ~83.51(28) & ~83.57(2) & ~27 &	~16 &
{}~7 & 9 & 17\\
$\Gamma_h$ (MeV) & ~1740(6) & ~1742(5) & -1 & -3 &
-6  & 1.5 & 3\\
$\Gamma_Z$ (MeV) & ~2487(7) & ~2490(5) & ~3.6 & ~0.5 &
-3.6 & 2.1 & 4.2\\
$\sigma_h$ (nb) & ~41.45(17) & ~41.44($\underline{5}$) & -0.4 & -0.4 &
-0.4 & 0.7 &
0.11\\
$R_{\ell}\equiv\Gamma_h/\Gamma_{\ell}$ & ~20.83(6) & ~20.84(6) & -6 &
-7.4 & -8.2 & $\underline{0.3}$ & $\underline{0.6}$\\
$R_b\equiv\Gamma_b/\Gamma_h$ & ~0.2201(32) & ~0.2197($\underline{1}$) &
-31 & -31 & -31 & $\underline{3}$ & $\underline{7}$\\
\hline \end{tabular}
\end{center}
\end{table}

\noindent
{\bf Table - } Comparison of experimental values of various LEP
observables
with the electroweak ``Born" approximation (column 3). The quotation
marks indicate that for the hadronic decays the virtual gluons are
taken into account by the universal factor $G$ [see Eq. (33) and the
discussion following Eq. (35)].

The next three columns (4,5,6) give
the values of the MSM one-loop electroweak corrections to the ``Born"
approximation (and also of specific gluonic corrections depending on
$m_b$ and $m_t$) for $m_t$ = 150 GeV and for three values of
$m_H$: 100, 300 and 1000 GeV, respectively.
The last two columns, 7 and 8, show the increments of the MSM
corrections for
$\Delta m_t = \pm 10$ GeV and $\pm$ 20 GeV, respectively. As not all
terms of
the order $\bar\alpha\bar\alpha_s$ have been taken into account, the
numbers in columns 4-8
should not be taken too litterally.

The $1\sigma$
uncertainties in columns 2 and 3 are quoted in brackets, with the
customary convention regarding digits. The figures in columns 4-8 are
again given in
the corresponding units.

The Born uncertainties for $m_W/m_Z$ and $g_V/g_A$ derive mainly from
$\Delta\bar\alpha /\bar\alpha = \pm 9.3\times 10^{-4}$. The
uncertainty of $g_A$ is ``hidden" in $\Gamma_0$ [Eq. (22)]. The
uncertainties of hadronic observables in column 3 corresopnd to
$\Delta\bar\alpha_s = \pm 0.01$.
A figure for uncertainty or shift is underlined when the coefficient in
front of
$\Delta\bar\alpha_s$ or $\Delta m_t$ is negative.

\vfill\eject

\vspace*{1cm}
\begin{center}
{\bf FIGURE CAPTIONS}
\end{center}

\begin{description}
\item[Fig. 1] $V_R$  the one-loop radiative correction to the ratio $R =
g_V/g_A$ defined by Eqs. (20) and (40). The curves 1, 2, 3 correspond to
$m_H$ =
50, 300 and 1000 GeV, respectively. The solid horizontal line
corresponds to the
central experimental value $\bar V_R$ and the dashed lines to $\pm\delta
V_R$
as given by Eq. (45). The dotted parabola describes the
$m^2_t$-dependence,
which dominates at large values of $m_t$. These curves are taken from
Ref. [4],
which contains similar graphs for $V_m$ and $V_A$; the experimental
value of
$V_R$ is a new one.
\item[Fig. 2] Isolines corresponding to the central values
$\bar V_A, \bar V_m$ and $\bar V_R$ (solid lines) and to $V_A + \delta
V_A, \bar
V_m-\delta V_m, \bar V_R - \delta V_R$ (dashed lines). The horizontal
dashed
line at $m_H$ = 700 GeV shows the theoretical upper limit for the mass
of an
elementary Higgs. The horizontal dashed line at $m_H$ = 62 GeV shows the
lower
experimental bound for $m_H$ from direct searches at LEP.  The figure is
taken
from Ref. [36].
\end{description}


\begin{thebibliography}{99}
\bibitem{aaa} Review of Particle Properties, \PR {\bf D15} N11, Part II
(1992),
p. III.59.
\bibitem{bb}  C. DeClercq, Talk given at XXVIIIth Rencontres de Moriond
``Electroweak Interactions and Unified Theories", Les Arcs, 1993.
\bibitem{cc} V. Innocente, Talk given at XXVIIIth  Rencontres de Moriond
``Electroweak Interactions and Unified Theories", Les Arcs, 1993.
\bibitem{dd} M. Pepe Altarelli, Talk given at Les Rencontres de Physique
de la
Vall\'ee d'Aoste ``Results and Perspectives in Particle PHysics, La
Thuile,
1993.
\bibitem{ee} R. Tenchini, Talk given at XXVIIIth Rencontres de Moriond
``Electroweak Interactions and Unified Theories", Les Arcs, 1993.
\bibitem{ff} V.A. Novikov, L.B. Okun and M.I. Vysotsky, CERN Preprint
TH. 6053/91
(1991); TPIMINN-91/14-T (1991).
\bibitem{ggg} V.A. Novikov, L.B. Okun and M.I. Vysotsky, \NP {\bf B397}
(1993)
35.
\bibitem{hh} V.A. Novikov, L.B. Okun and M.I. Vysotsky, CERN Preprint
TH.
6855/93 (1993)
\bibitem{jj} F. Jegerlehner, Villigen Preprint PSI-PR-991-08 (1991).
\bibitem{kk} H. Burkhardt, F. Jegerlehner, G. Renso and C. Verzegnassi,
\ZP
{\bf C43} (1989) 497.
\bibitem{lll} S.L. Glashow, \NP {\bf 22} (1961) 579.
\bibitem{mm} S. Weinberg, \PRL {\bf 19} (1967) 1264.
\bibitem{nn} A. Salam, in Elementary Particle Theory, ed. N. Svartholm
(Almquist and Wiksells, Stockholm, 1969), p. 367.
\bibitem{oo} L.B. Okun, Leptons and Quarks (North Holland, Amsterdam,
1982).
\bibitem{pp} A. Sirlin, \PR {\bf D22} (1980) 971.
\bibitem{qq} L. Rolandi, Plenary Talk at XXVIth International Conference
on
High Energy Physics, Dallas, CERN Preprint PPE/92-175 (1992).
\bibitem{rr} S. Gorishny, A. Kataev and S. Larin, \PL {\bf B259} (1991)
144.
\bibitem{ss} L. Surguladze and M. Samuel, \PRL {\bf 66} (1991) 560.
\bibitem{tt} A. Kataev, \PL {\bf B287} (1992) 209.
\bibitem{uu} G. Altarelli, Rapporteur Talk at the Conference ``QCD-20
Years
Later", Aachen, 1992, CERN Preprint TH. 6623/92 (1992).
\bibitem{vv} S. Bethke, Plenary Talk at XXVIth International Conference
on
High Energy Physics, Dallas, 1992. Preprint HD-PY 92/13, OPAL-CP093
(1993).
\bibitem{ww} B. Kniehl and J. K\"uhn, \NP {\bf B329} (1990) 547.
\bibitem{yy} B. Kniehl and J. K\"uhn, \NP {\bf B224} (1989) 229.
\bibitem{zz} J. K\"uhn, in Proceedings XXVth International Conference on
High
Energy Physics, Singapore, 1990, eds. K.K. Phua and Y. Yamaguchi (World
Scientific, Singapore, 1991), Vol. II, p. 1451.
\bibitem{aai} K. Chetyrkin and J. K\"uhn, \PL {\bf B248} (1990) 359.
\bibitem{bbi} A. Kataev and V. Kim, Preprint ENSLAPP-A-407/92 (1992).
\bibitem{cci} A. Kataev, Talk given at Quark '92 Conference, Zvenigorod,
to be
published in ``Quark '92", Singapore, 1993.
\bibitem{ddi}CHARM II Collaboration, P. Vilain et al., \PL {\bf B281}
(1992)
159.
\bibitem{eei} V.A. Novikov, L.B. Okun and M.I. Vysotsky, \PL {\bf B299}
(1993)
329; E {\bf B304} (1993) 386.
\bibitem{ffi} V.A. Novikov, L.B. Okun and M.I. Vysotsky, \PL {\bf B298}
(1993)
453.
\bibitem{ggi} G. Altarelli, R. Kleiss and C. Verzegnassi,
ets., Physics at LEP~1, Vol. 1, CERN Report 89-08 (CERN, Geneva, 1989).
\bibitem{hhi} G. Altarelli, CERN Preprint TH. 6867/93 (1993).
\bibitem{jji} J. Ellis, G. Fogli and E. Lisi, \PL {\bf B292} (1992) 427.
\bibitem{kki} N. Nekrasov, V.A. Novikov and M. Vysotsky, CERN Preprint
TH.
6696/92 (1992), to be published in Yadernaya Fizika.
\bibitem{lli}  G. Altarelli, R. Barbieri and F. Caravaglios, CERN
Preprint TH.
6770/93/Rev. (1993).
\bibitem{mmi} V.A. Novikov, L.B. Okun, M.I. Vysotsky and V.P. Yurov CERN
Preprint
TH. 6849/93 (1993).
\end{thebibliography}
\end{document}